\documentclass[aps,pra,twocolumn,superscriptaddress,showpacs,longbibliography]{revtex4-1}

\usepackage{CJK}
\usepackage{graphicx}
\usepackage{bm}
\usepackage{amssymb}
\usepackage{amsmath}
\usepackage[colorlinks,linkcolor=blue,anchorcolor=blue,citecolor=blue]{hyperref}
\usepackage{xcolor}

\begin{document}

\title{Current-driven magnetic resistance in van der Waals spin-filter antiferromagnetic tunnel junctions with MnBi$_2$Te$_4$}

\author{Lishu Zhang}
\email{lis.zhang@fz-juelich.de}
\affiliation{Peter Gr{\"u}nberg Institut (PGI-1) and Institute for Advanced Simulation (IAS-1), Forschungszentrum J{\"u}lich, J{\"u}lich 52428, Germany}

\author{Hui Li}
\affiliation{Key Laboratory for Liquid-Solid Structural Evolution and Processing of Materials, Ministry of Education, Shandong University, Jinan 250061, China}

\author{Yanyan Jiang}
\affiliation{Key Laboratory for Liquid-Solid Structural Evolution and Processing of Materials, Ministry of Education, Shandong University, Jinan 250061, China}

\author{Zishen Wang}
\affiliation{Department of Physics, National University of Singapore, Singapore 117542, Singapore}

\author{Tao Li}
\affiliation{Department of Physics, Hong Kong University of Science and Technology, Clear Water Bay, Kowloon, Hong Kong 999077, China}

\author{Sumit Ghosh}
\affiliation{Peter Gr{\"u}nberg Institut (PGI-1) and Institute for Advanced Simulation (IAS-1), Forschungszentrum J{\"u}lich, J{\"u}lich 52428, Germany}
\affiliation{Institute of Physics, Johannes Gutenberg-University Mainz, 55128 Mainz, Germany}

\begin{abstract}
The field of 2D magnetic materials has paved the way for the development of spintronics and nanodevices with new functionalities. Utilizing antiferromagnetic materials, in addition to layered van der Waals (vdW) ferromagnetic materials, has garnered significant interest.
In this work, we present a theoretical investigation of the behavior of MnBi$_2$Te$_4$  devices based on the non-equilibrium Green's function method. Our results show that the current-voltage (I-V) characteristics can be influenced significantly by controlling the length of the device and bias voltage and thus allow us to manipulate the tunneling magneto-resistance (TMR) with an external bias voltage. This can be further influenced by the presence of the boron nitride layer which shows significantly enhanced TMR by selectively suppressing specific spin channels for different magnetic configurations. By exploiting this mechanism, the observed TMR value reaches up to 3690\%, which can be attributed to the spin-polarized transmission channel and the projected local density of states. Our findings on the influence of structural and magnetic configurations on the spin-polarized transport properties and TMR ratios give the potential implementation of antiferromagnetic vdW layered materials in ultrathin spintronics.

\end{abstract}

\maketitle

\section{Introduction}

In recent years, there has been a growing interest in exploring the potential of antiferromagnetic materials for spintronics applications.\cite{han2023coherent} Antiferromagnetism is a type of magnetic order in which the spins of adjacent atoms align in opposite directions, resulting in a net magnetic moment of zero. Antiferromagnetic materials have several advantages over their ferromagnetic counterparts, including faster spin dynamics,\cite{xu2022exchange, yan2022quantum} lower susceptibility to external magnetic fields,\cite{wu2022antiferromagnetic} and absence of stray fields that can interfere with nearby devices\cite{achilli2022magnetic}.

Antiferromagnetic spintronics is a relatively new field that aims to exploit the unique properties of antiferromagnetic materials for spin-based information processing and storage. One of the key challenges in antiferromagnetic spintronics is the development of efficient ways to manipulate the spin and charge transport in these materials \cite{lima2021antiferromagnetic, stepanov2022tuning}. One promising approach is the use of antiferromagnetic tunnel junctions (AFM-TJs), which consist of two electrodes separated by a thin layer of antiferromagnetic material. By applying an external bias voltage, it is possible to control the spin-dependent transport properties of the junction and achieve a high degree of spin polarization. AFM-TJs have several advantages over the traditional magnetic tunnel junctions (MTJs)\cite{zhang2021recent}, such as the absence of stray fields and the possibility of achieving a high tunnel magnetoresistance (TMR) ratio even at room temperature.\cite{dong2022tunneling} Moreover, AFM-TJs offer the possibility of using antiferromagnetic materials with high Neel temperature, such as MnB$_2$Te$_4$(MBT), which is a compound that has attracted a lot of attention in recent years due to its unique properties.

The TMR effect is the basis of various spintronic devices, for instance, magnetoresistive random access memories (MRAM),\cite{ney2003programmable, wang2018current, wu2021magnetic} magnetic field sensors,\cite{li2021spin}\cite{kim2019tailored} racetrack memory,\cite{raymenants2021nanoscale, cao2020nonvolatile} and spin logics,\cite{huang2021high, lin2019two} which have shown huge potentials in the post-Moore era. To govern TMR performance, spin polarization is the most important factor for TJs because a highly spin-polarized current is essential for high magnetoresistance. Various approaches have been explored to improve spin polarization, for example, using MoS$_2$,\cite{zhang2016magnetoresistance, galbiati2019path} graphene or boron nitride (BN) as barrier layer,\cite{liu2016exfoliating, li2019spin} half-metals as electrodes\cite{reza2019topological}. Based on theoretical predictions, spin polarization can be improved using these methods, by overcoming interface disorder and allowing the Fermi level crossing only one spin channel. However, actual device performance has been far below expectations. Thus, the usage of pinning layers to regulate the spin polarization is worth another attempt.
Moreover, because of reduced dimensionality and novel physical properties, 2D materials are expected to provide a reliable solution to the problems in the manufacturing of high-performance TJs through the layer-by-layer control of the thickness, sharp interfaces, and high perpendicular magnetic anisotropy (PMA).\cite{zhang2021recent} A number of 2D antiferromagnets have been discovered and their potential in spintronic devices has been demonstrated. Among them, layered vdW MnBi$_2$Te$_4$ (MBT) has been successfully synthesized recently and confirmed experimentally to be an intrinsic magnetic topological insulator,\cite{deng2020quantum}\cite{li2019intrinsic} which caught immediate attention in the fields of topology\cite{liu2021magnetic}\cite{wu2020toward} and spintronics.\cite{zhang2019experimental} Recent experimental studies have demonstrated the feasibility of fabricating MBT-based tunnel junctions and measuring their transport properties. Theoretical studies have also been conducted to shed light on the underlying mechanisms that govern the spin and charge transport in these materials. For example, its application in MTJ has been reported recently by Yan et al.\cite{yan2021barrier} Specifically, MBT MTJs with BN, graphene, and vacuum as tunnel barrier and graphene electrode have been investigated. While these studies demonstrate the potential of MBT in MTJs, it is noted that the performance of the MTJ could be further improved by fine-tuning the structure and parameters. For example, employing a pinning layer and tuning the barrier thickness might be beneficial for enhancing the spin polarization and TMR \cite{yan2020significant}.

Motivated by the above, we design a series of MBT-based AFM-TJ devices, including an even number of MBT layers with or without intermediate BN layer(s), and a monolayer case as the benchmark. Using density functional theory (DFT) calculations combined with nonequilibrium Green’s function (NEGF) technique, the origin of spin-polarized transport through single-layer, double-layer, and four-layer MBT sandwiched between metal electrodes are investigated. The influence of the number of hexagonal BN layers on transport properties and TMR ratio is studied. Driven by voltage, a high-quality magnetic tunnel junction with remarkable magnetoresistance is realized in ultrathin MBT-BN-MBT-MBT-MBT multilayer structures. The study demonstrates the feasibility of 2D vdW layered antiferromagnets in ultrathin spintronic devices and their applications in spintronics.

\section{Methods}

The electron transport properties were investigated using the DFT in combination with the NEGF method\cite{brandbyge2002density, taylor2001ab, soler2002siesta} implemented in the QuantumATK package.\cite{ni2013transport, zhang2019taper} The Perdew-Burke-Ernzerhof (PBE) formulation of the generalized gradient approximation (GGA)\cite{perdew1996generalized} is used for the exchange-correlation functional. The double-zeta plus polarization (DZP) basis set is adopted for all atoms. The mesh cut-off used for the electrostatic potentials is 85 Ha, and the temperature in the Fermi function is set to 300 K. The k-point set for devices in the ATK package is 4(n$_a$), 4(n$_b$), and 100(n$_c$), respectively, where direction \textit{c} is transport direction. The average Fermi level which is the average chemical potentials of the left and the right electrodes is set to zero. Moreover, van der Waals force was considered through the DFT-D2 method \cite{grimme2006semiempirical} in all the structural optimizations.
The current passing through the device is calculated by the Landauer-like formula, \cite{meir1992landauer,datta1997electronic} which is expressed as
\begin{equation}
I = \frac{2e}{h} \int_{\mu_R}^{\mu _L} T(E) dE
\end{equation}
where  $\mu_R$ and  $\mu_L$ are the chemical potentials of the right and left electrodes respectively which are chosen symmetrically such that $\mu_L$ = -$\mu_R$ = eV/2, with \textit{V} being the bias voltage.  
The transmission spectrum is obtained from 
 \begin{equation}
 T(E) = \mathrm{Tr}[\Gamma_L(E)G^{R}(E)\Gamma_R(E)G^{A}(E)]
 \end{equation}
where $G^{R}(E)$  is the retarded Green’s functions and $G^{A}(E)$  is the advanced Green’s functions of the scattering region.   $\Gamma_{L,(R)}(E)= i(\Sigma_{L(R)}^R - \Sigma_{L(R)}^A)$ is a coupling function between the structure and the left (right) electrode, and  $\Sigma_{L(R)}^R(A)$ is a self-energy matrix for considering the influence of the left and right half-infinite electrodes.

\begin{figure*}
\centering
\includegraphics[width=\textwidth]{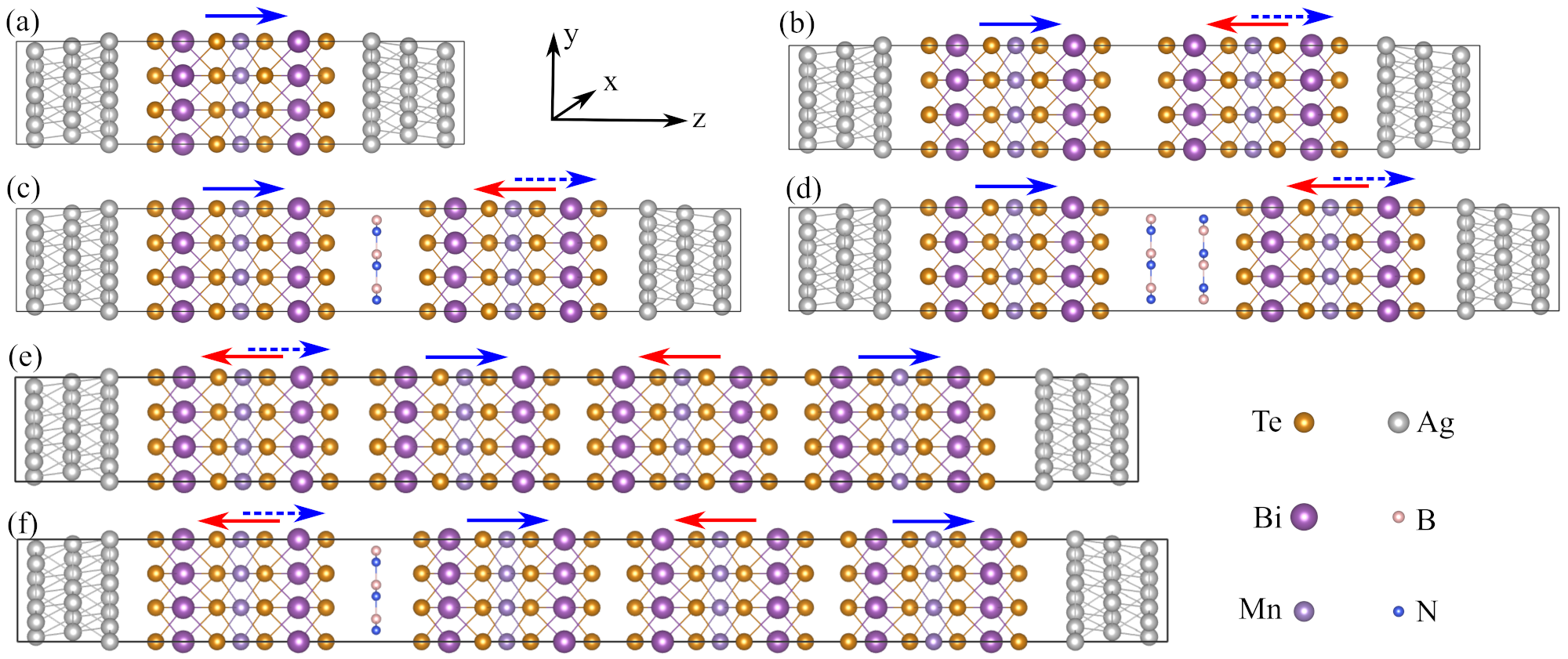}
\caption{Structures of MBT magnetic tunnel junction devices. (a) Ag-MBT-Ag, (b) Ag-MBT-MBT-Ag, (c) Ag-MBT-BN-MBT-Ag, (d) Ag-MBT-BN-BN-MBT-Ag, (e) Ag-MBT-MBT-MBT-MBT-Ag, and (f) Ag-MBT-BN-MBT-MBT-MBT-Ag, respectively. Blue and red arrows indicate the magnetic moment direction of Mn atoms. The layer marked with both blue and red arrows is a free layer. Solid lines show antiparallel (AP) magnetization configuration while the dashed line shows the parallel (P) configuration obtained by reversing the direction of magnetization of the free layer. These devices are periodic in \textit{x} and \textit{y} directions, and current flows in \textit{z} direction.}
\label{fig:1}
\end{figure*}

\begin{figure*}
\centering
\includegraphics[width=\textwidth]{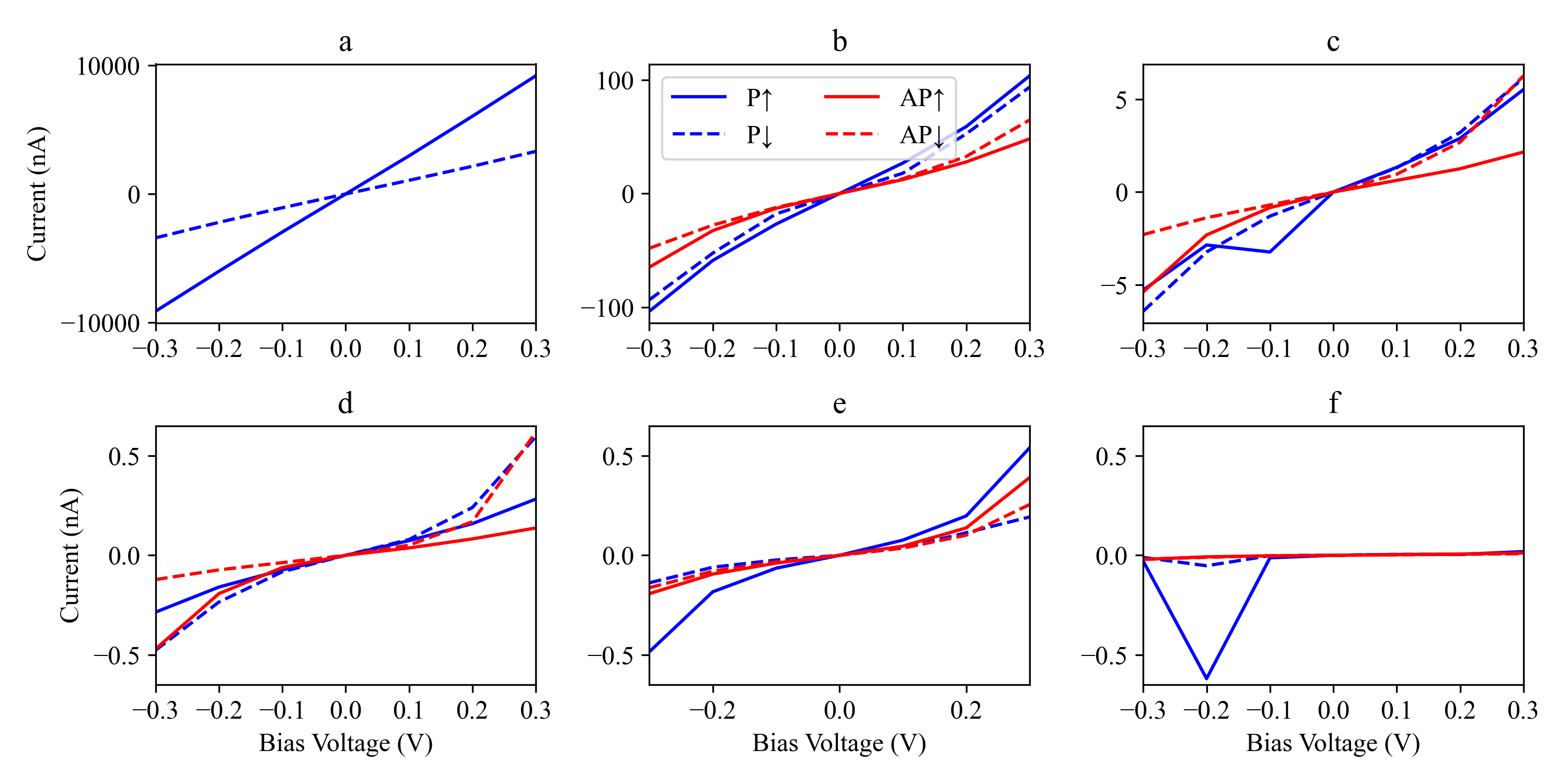}
\caption{I-V characteristic curves for spin up and spin down of (a) Ag-MBT-Ag, (b) Ag-MBT-MBT-Ag, (c) Ag-MBT-BN-MBT-Ag, (d) Ag-MBT-BN-BN-MBT-Ag, (e) Ag-MBT-MBT-MBT-MBT-Ag, and (f) Ag-MBT-BN-MBT-MBT-MBT-Ag, respectively. The red and blue lines show the P and AP configuration while the solid and dashed lines show the contribution from spin up and down channels.}
\label{fig:2}
\end{figure*}

\section{Results And Discussion}

The main objective of this work is to explore a device configuration that provides superior TMR. In this study, we consider six different device configurations (Fig. \ref{fig:1}) and compare their properties to find the most suitable configuration. The tunnel junctions contain an even number of MBT layers to ensure antiferromagnetic alignment with an optional BN layer in between.
Note that MBT is a well-known intrinsic magnetic topological insulator \cite{Li2019} which demonstrates antiferromagnetic nature in multilayer configuration. MBT mono layer possesses a band gap of $\sim$800 meV \cite{Trang2021} which can reduce to $\sim$80 meV for 5 layers. However, the low energy states are strongly localized at the surface states and conduct along the $xy$ plane. The topological states can be utilized for efficient switching the magnetization of the free layer. In our device configuration the current flows along the $z$ direction which has to go through the insulating bulk region. Therefore the transmission is mostly dominated by the tunnelling mechanism. For our study, a layer adjacent to the electrode is kept as the free layer where the magnetization can be manipulated by means of spin-orbit torque \cite{Liu2021b, GarciaOvalle2023, Tang2023}. This can change the magnetic configuration of the first pair of MBT from anti-parallel (AP) to parallel (P) configuration (Fig.\ref{fig:1}) \cite{Dolui2020b}. The transport properties and TMR is calculated with a two-terminal device configuration where 
these heterojunctions are placed in the central region, between two semi-infinite 3×3 $<111>$-cleaved surfaces of bulk Ag electrodes. The choice of silver electrodes presents several advantages due to their high conductivity, chemical stability, and low contact resistance, which result in improved device performance and reliability. Negligible strains are applied to silver (0.79 \%) and BN (0.37 \%), resulting in a lattice constant: a=b=7.55 Å. And lattice angle is $\alpha$=$\beta$=90$^\circ$, $\gamma$= 60$^\circ$.To create the interface, supercells of the two surfaces are aligned and matched by applying strain on the silver or BN surfaces. The use of large supercells helps minimize lattice mismatch between the two surfaces. Unlike previous works where high interface strains were observed, often exceeding 3\% and up to 5\% for certain metals \cite{yan2018monolayer}, in this study, we have enlarged the unit cells into supercells to ensure the interface strain remains small enough to be considered negligible.

\begin{figure}
\centering
\includegraphics[width=0.5\textwidth]{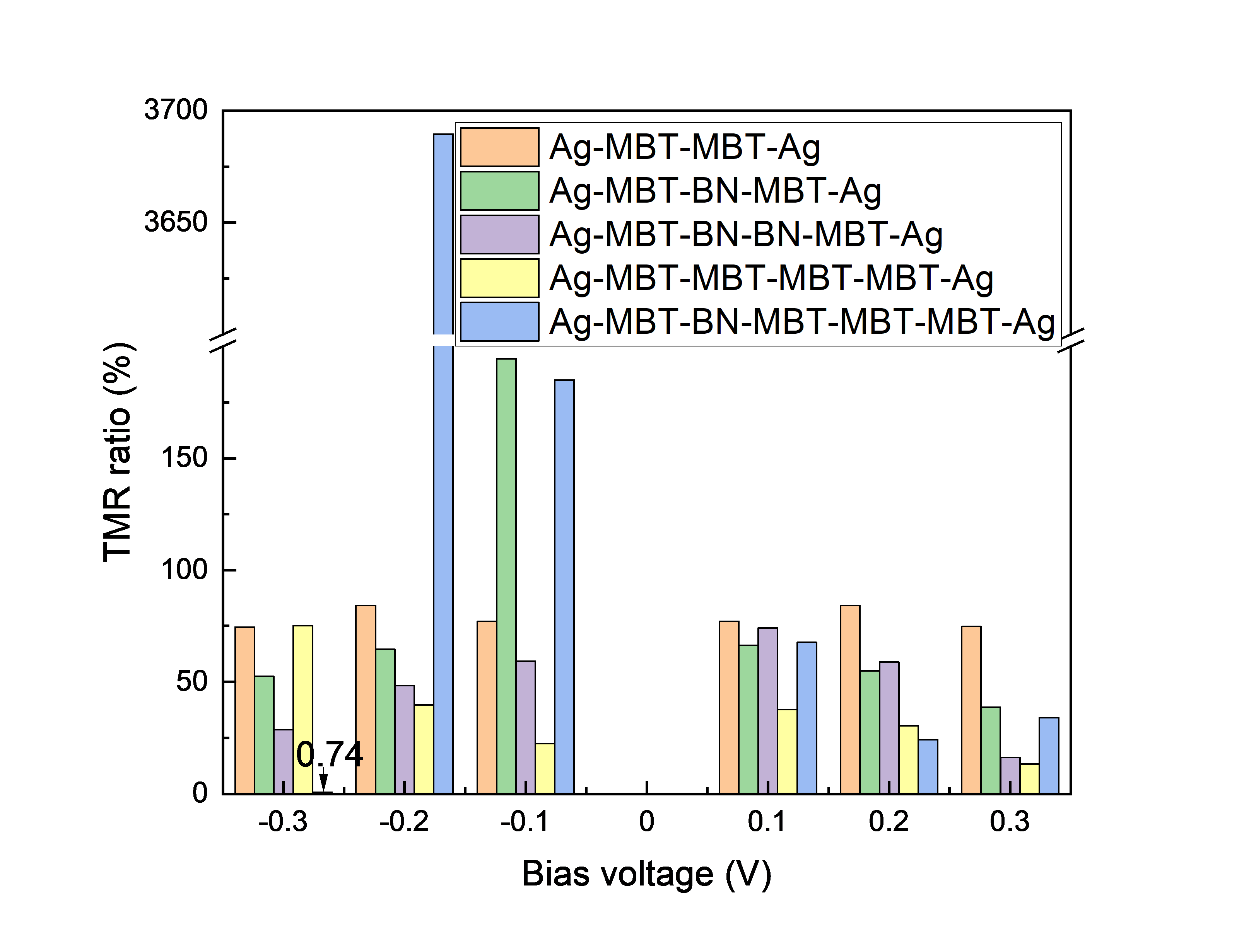}
\caption{TMR of (a) Ag-MBT-Ag, (b) Ag-MBT-MBT-Ag, (c) Ag-MBT-BN-MBT-Ag, (d) Ag-MBT-BN-BN-MBT-Ag, (e) Ag-MBT-MBT-MBT-MBT-Ag, and (f) Ag-MBT-BN-MBT-MBT-MBT-Ag, respectively. }
\label{fig:3}
\end{figure}

As mentioned before, here we consider six different configurations (Fig.\ref{fig:1}).  The first configuration is a mono-layer of MBT which is used to benchmark the transport properties of the MBT layer. Consecutive configuration contain an even number of MBT layers with or without intermediate BN layer(s).
We first calculated the I-V curves for the P and AP magnetization configurations of the six AFM-TJ devices for bias voltage V ranging from $-0.3$ V to $0.3$ V (Fig.\ref{fig:2}) and derive the tunnel current and the TMR ratio of the devices (Fig.\ref{fig:3}). The TMR ratio is given as $TMR=\left(R_{\mathrm{AP}}-R_{\mathrm{P}}\right) / R_{\mathrm{P}} \times 100 \%=\left(I_{\mathrm{P}}-I_{\mathrm{AP}}\right) / I_{\mathrm{AP}} \times 100 \%$, where $R_{\mathrm{P}}$ and $R_{\mathrm{AP}}$ are the resistance under the P and AP magnetic configurations, respectively.

For a systematic study, first we consider the monolayer MBT (Ag-MBT-Ag, Fig.\ref{fig:1}a). Its I-V curve shows a linear behavior as expected (Fig.\ref{fig:2}a). The spin-up channel shows a higher slope due to its higher occupation. For two layers of MBT (Ag-MBT-MBT-Ag, Fig.\ref{fig:1}b), the current decreases by a factor of 100 due to the inter-layer tunneling resistance (Fig.\ref{fig:2}b). Note that for AP configuration the layers have opposite spin which preserves the time-reversal symmetry and a reflection symmetry where the mirror plane lies in between two MBT layers. This results in an anti-symmetric current from each spin channel such that ($I_{\uparrow,\downarrow}^{AP}(+V) = -I_{\downarrow,\uparrow}^{AP}(-V)$). The currents for the P configuration on the other hand follow $I_{\uparrow,\downarrow}^{P}(+V) = -I_{\uparrow,\downarrow}^{P}(-V)$. The symmetry is further reduced with an increase in tunneling resistance if a BN layer is introduced in between MBT layers (Ag-MBT-BN-MBT-Ag, Fig.\ref{fig:1}c). The I-V characteristics become more asymmetric in this case. Note that, for P configuration the BN layer can provide resonant tunneling which causes a slight enhancement in the current for a bias voltage of -0.1V. Introducing a second layer of BN with AA' \cite{Gilbert2019, Yasuda2021} stacking (Ag-MBT-BN-BN-MBT-Ag, Fig.\ref{fig:1}d) can restore the symmetry resulting an antisymmetric I-V characteristic ($I_{\uparrow,\downarrow}^{AP}(+V) \approx -I_{\downarrow,\uparrow}^{AP}(-V)$, $I_{\uparrow,\downarrow}^{P}(+V) \approx -I_{\uparrow,\downarrow}^{P}(-V)$, Fig.\ref{fig:2}d). The situation becomes more complicated if we consider a device with 4 MBT layers (Ag-MBT-MBT-MBT-MBT-Ag, Fig.\ref{fig:1}e), and the decrease of current due to multiple tunnel junction becomes more prominent. Note that two layers of BN in between two layers of MBT result in the same magnitude of current as the four layers of MBT. One can further enhance the asymmetry of the structure by introducing a BN layer (Ag-MBT-BN-MBT-MBT-MBT-Ag, Fig.\ref{fig:1}f). Inclusion of a single BN layer can also enhance the transmission of one spin while suppressing the other channel (Fig.\ref{fig:2}f) which has been also observed in Ag-MBT-BN-MBT-Ag configuration (Fig.\ref{fig:1}c). Although the length of the device along with its heterogeneous structure suppresses the current in both channels significantly, the resonant transmission produces a comparatively high value of the current due to the spin-up channel for $V_{Bias}$=-0.2V resulting in an exceptionally high value of TMR (Fig.\ref{fig:3}).

\begin{figure}[hb!]
\centering
\includegraphics[width=\linewidth]{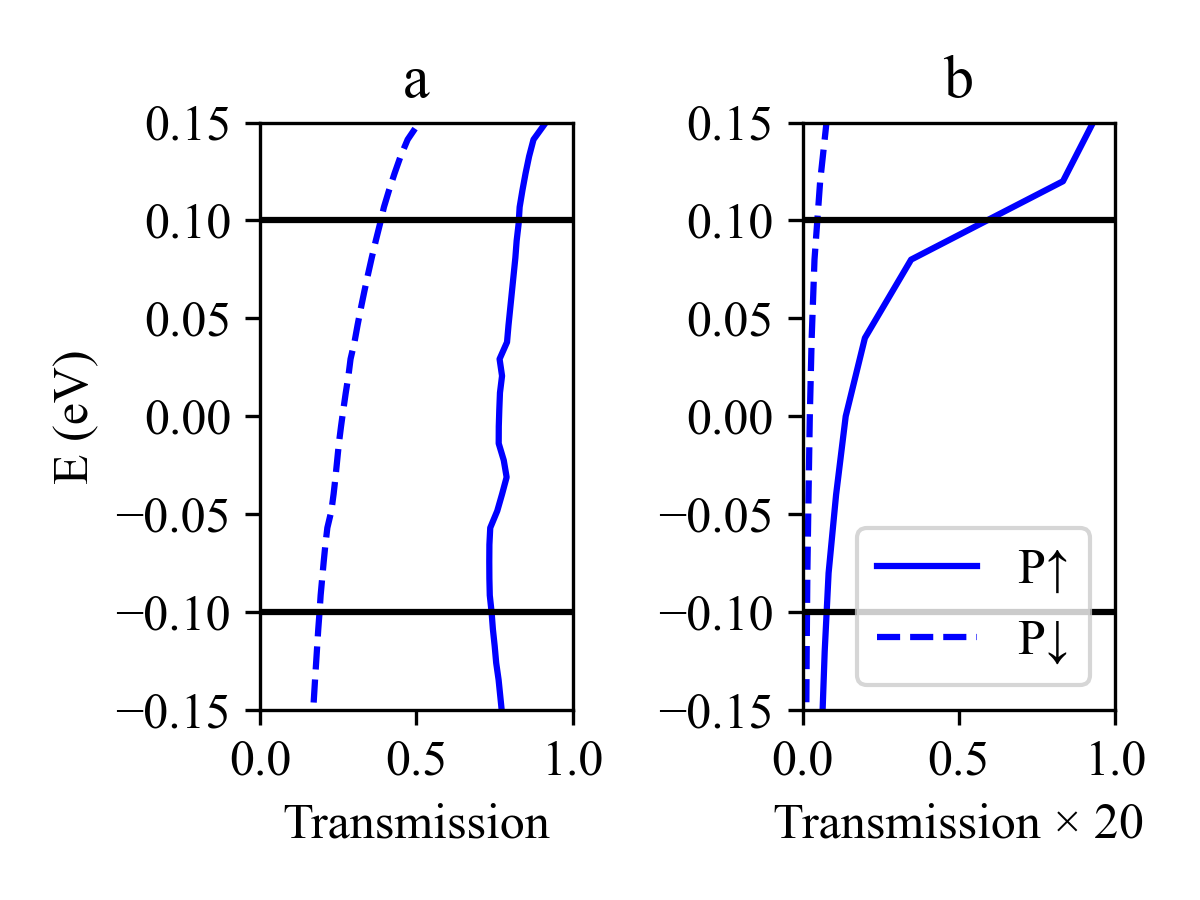}
\caption{Transmission spectra for (a) Ag-MBT-Ag and (b) Ag-MBT-MBT-Ag under 0.2V bias. Solid and dashed lines show the contribution from up and down spin channel. Horizontal black lines show the bias window.}
\label{fig:4}
\end{figure}

To understand the impact of the tunneling resistance, first we compare the Ag-MBT-Ag and Ag-2MBT-Ag layers. We choose an intermediate bias voltage of 0.2V and consider the ferromagnetic or P configuration (Fig.\ref{fig:4}). From Fig.\ref{fig:4}, one can see that the magnitude of the transmission spectra decreases by a factor of 20 due to the presence of the second layer. For a better understanding, we look at the different transmission channels coming from the Ag electrode. At $E$=0eV there are a total of six channels that contribute to the transport (Table.\ref{tab:1}). The channel that corresponds to the highest transmission occurs for the down spin channel for Ag-MBT-Ag configuration (Channel 1 MBT($\downarrow$)). We scale the rest of the transmission eigenvalues with respect to this value to analyze their relative contribution. Note that although the highest transmission occurs for the down spin channel, only two out of six channels contribute significantly. For the spin-up channel on the other hand all the channels contribute with almost equal weight which makes the total contribution from the spin-up channel higher than the spin-down channel. Similar behavior can be observed for Ag-MBt-MBT-Ag configuration with an average two orders of magnitude smaller contribution.

\begin{table}[htbp]
\centering
\begin{ruledtabular}
\begin{tabular}{c|cccc}
Channel & MBT($\uparrow$) & MBT($\downarrow$) & 2MBT($\uparrow$) & 2MBT($\downarrow$) \\ \hline
1 & 0.65 & 1.00 & $8.37 \times 10^{-3}$ & $4.38 \times 10^{-3}$ \\ 
2 & 0.55 & 0.99 & $8.37 \times 10^{-3}$ & $4.38 \times 10^{-3}$ \\
3 & 0.55 & $1.24 \times 10^{-2}$ & $4.99 \times 10^{-4}$ & $1.84 \times 10^{-6}$ \\
4 & 0.47 & $8.14 \times 10^{-3}$ & $2.59 \times 10^{-4}$ & $3.70 \times 10^{-7}$ \\
5 & 0.42 & $7.91 \times 10^{-3}$ & $2.25 \times 10^{-4}$ & $3.32 \times 10^{-7}$ \\
6 & 0.42 & $6.10 \times 10^{-3}$ & $2.25 \times 10^{-4}$ & $3.23 \times 10^{-7}$ \\ \hline
Total & 3.07 & 2.03 & $1.79 \times 10^{-2}$ & $8.76 \times 10^{-3}$
\end{tabular}
\end{ruledtabular}
\caption{Magnitude of transmission eigenvalues of six transport channels at $E$=0eV for P configurations of Ag-MBT-Ag and Ag-MBT-MBT-Ag with 0.2V bias voltage. The values are scaled with respect to the maximum transmission eigenvalue 0.26 which occurs at the first channel for spin down of MBT configuration.}
\label{tab:1}
\end{table}

\begin{figure}[h]
\centering
\includegraphics[width=\linewidth]{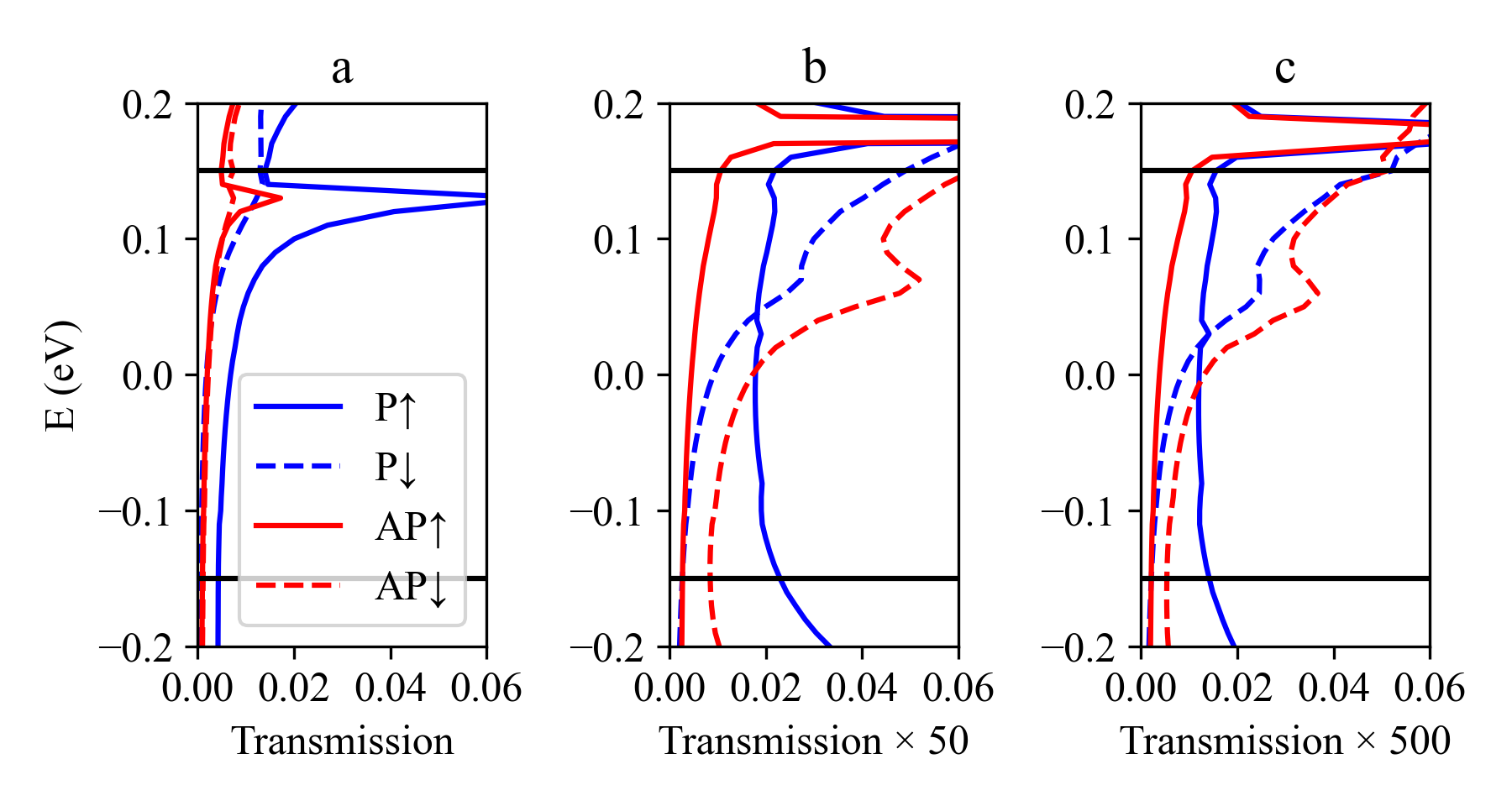}
\caption{Transmission spectra and spin difference density of (a) Ag-MBT-MBT-Ag, (b) Ag-MBT-BN-MBT-Ag, and (c) Ag-MBT-BN-BN-MBT-Ag under 0.3V bias. Blue and Red lines show the P and AP configuration while solid and dashed lines show the contribution from up and down spin channel. Horizontal black lines denote the bias window.}
\label{fig:5}
\end{figure}

Inclusion of an additional BN layer can further enhance the suppression of different spin channels. To understand that, first, we consider three representative cases shown in Fig.\ref{fig:1}bcd, all consisting of two layers of MBT and 0,1,2 layers of BN respectively. Note that except for Ag-MBT-BN-MBT-Ag, both Ag-MBT-MBT-Ag and Ag-MBT-BN-BN-MBT-Ag have antisymmetric I-V characteristics. We choose the bias voltage 0.3V for our analysis since the differences in their features are more clear for higher bias voltage. From Fig.\ref{fig:4} one can readily see that the dominant contribution is coming from the states around energy 0.12eV which is the maximum for spin-up. For spin down the magnitude is substantially reduced which is expected since the electron is injected from the left electrode which first faces the MBT layer with up spin. For an AP configuration, the current has to pass through MBT layers with alternative magnetization which reduce the transmission for both channel substantially. One can still see the characteristic peak around $E \sim$0.12eV which ensures that in both cases the transmission is happening through the same tunneling state. The presence of BN layers increases the local charge density which pushes this peak higher in energy (Fig.\ref{fig:5}b,c). As a result for the same bias voltage, the transmission decreases significantly.

\begin{figure}[h]
\centering
\includegraphics[width=\linewidth]{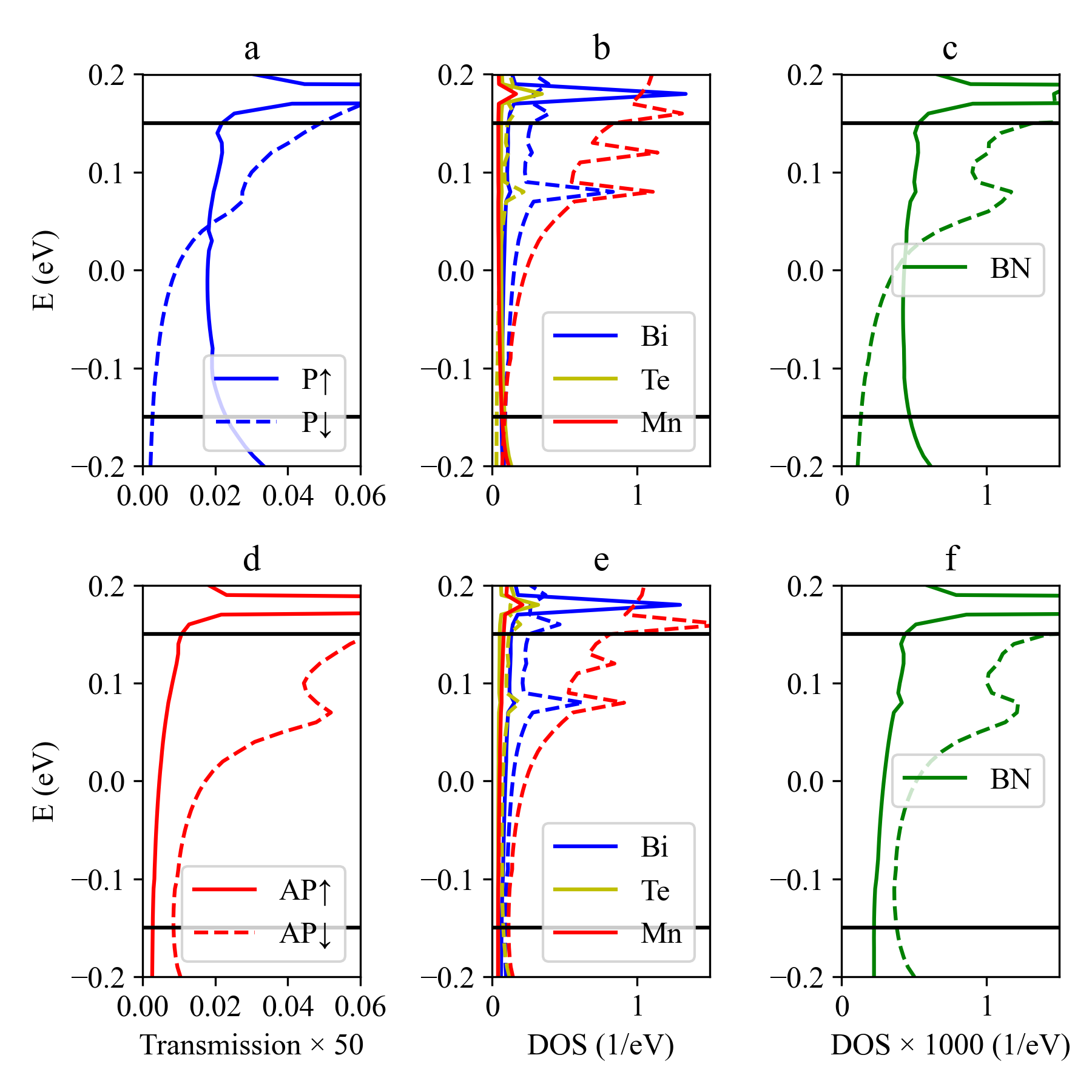}
\caption{Transmission spectra (a,d) and projected DOS per atom (b,c,e,f) of Ag-MBT-BN-MBT-Ag configuration. a,b,c show the P configuration, and d,e,f show the AP configuration. Solid and dashed lines show the contribution from the up and down spin channel. Horizontal black lines show the bias window.}
\label{fig:6}
\end{figure}

\begin{figure*}[ht!]
\centering
\includegraphics[width=0.33\textwidth]{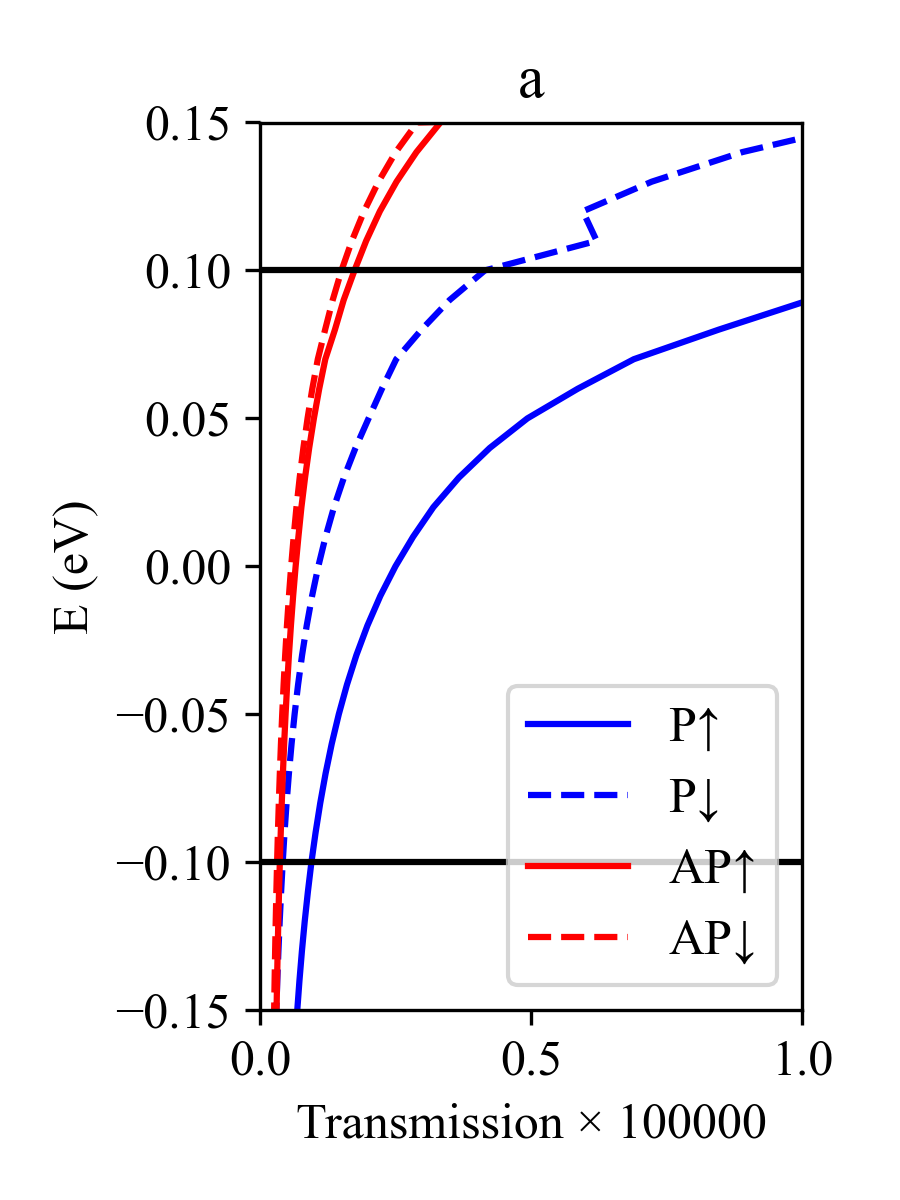}
\includegraphics[width=0.66\textwidth]{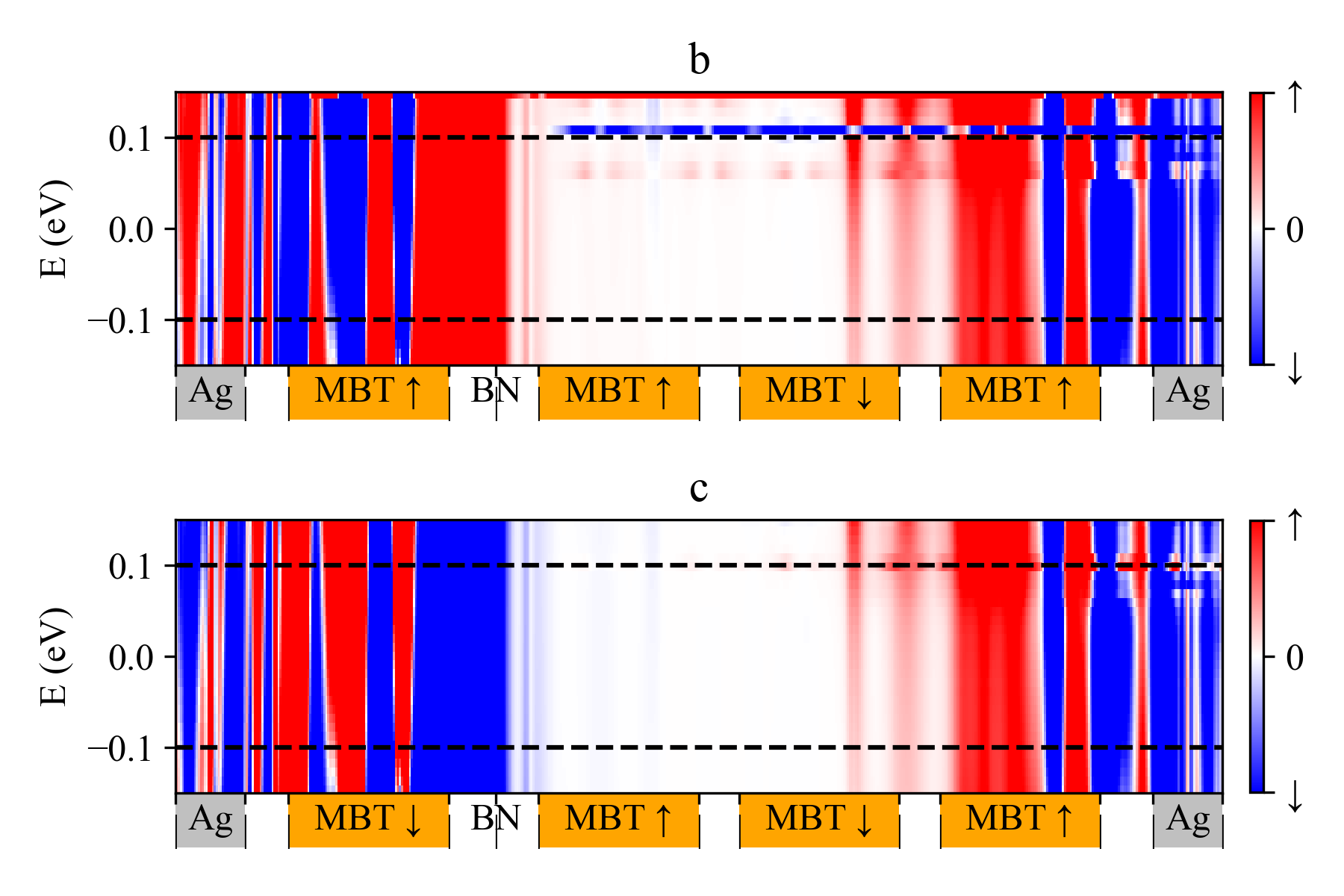}
\caption{The (a)transmission spectra and spin projected local density of states (PLDOS) of the Ag-MBT-BN-MBT-MBT-MBT-Ag configuration in the (b)P and (c)AP configuration under $-$0.2V. In b,c red and blue color correspond to up and down spin which is scaled with a maximum value $10^{-4}$eV$^{-1}$ to make the spin channels in the central region visible. The $x$ axis corresponds to the transport direction where the location of individual layers is shown with their magnetization profile. Horizontal black lines show the bias window.}
\label{fig:7}
\end{figure*}

Insertion of additional BN layer therefore plays a crucial role in controlling the current through different spin channels for P and AP configurations which results in non-trivial values of TMR (Fig.\ref{fig:3}).  Note that the presence of BN gives rise to a smaller peak in the bias window. To understand the origin of this peak we study the non-equilibrium projected DOS of the system (Fig.\ref{fig:6}) for Ag-MBT-BN-MBT-Ag configuration. Note that for both P and AP configurations, the down spin transmission channel has a stronger contribution at higher energy (Fig.\ref{fig:5}c, Fig.\ref{fig:6}a,d). For P configuration this might seem counterintuitive since both MBT layers up spin polarised. Such behavior can be caused by the finite interference effect and, as one can see from Fig.\ref{fig:2}e,f, does not take place for longer configurations. Similar behavior is observed in PDOS which is dominated by spin-down states. Note that, for Bi, Te, and Mn there are multiple peaks that do not correspond to a peak in the transmission spectrum. These peaks originate from the localized states which do not take part in the conduction. The PDOS of the BN on the other hand shows a peak exactly at the same energy where the transmission peak is. Since the BN layer is in the middle of the device, this peak is free from the strong localization that takes place at the layers adjacent to the electrode and corresponds to the extended states that take part in transmission.

Presence of multiple BN layers and an additional BN can, therefore, significantly influence the transmission through different spin channels. From the I-V characteristics, we have also seen that the transmission can be further influenced by the symmetry of the magnetic configuration as well. To receive the maximum benefit from all these criteria we finally consider an Ag-MBT-BN-MBT-MBT-MBT-Ag configuration at -0.2V bias which shows the highest TMR (Fig.\ref{fig:3}). Owing to its length and the presence of the additional BN layers, this configuration can facilitate a very small amount of current which is expected. From its transmission spectrum within the bias window, one can see that for both P and AP configurations, the transmission increases almost monotonically (Fig.\ref{fig:7}a). The long-range interference effect plays a dominant role which smears out any small local peaks. In spite of that, as one can see from Fig.\ref{fig:2}f and Fig.\ref{fig:7}a, for P configuration the up spin channel carries orders of magnitude more current than the down spin channel while the transmission for both spin channels in AP configuration is vanishingly small. To understand the origin of this behavior we calculate the LDOS for spin up and spin down separately and define spin projected LDOS (SPLDOS) as $\rho_S = \rho_\uparrow - \rho_\downarrow$. From Fig.\ref{fig:7}b one can see that for P configuration the central region is spanned by up spin channel only within the bias window. This makes the current due to the up spin significantly large. Physically one can attribute this behaviour to the configuration of magnetic tunnel junction. Note that for P configuration, the first two layers are parallel to each other which favors the passage of the corresponding spin channel. This is not possible for the AP configuration due to the anti-ferromagnetic alignment which significantly reduces the current due to both spins equally which makes the central region almost insulating (Fig.\ref{fig:7}c). Consequently one can obtain a significantly large total current for the P configuration ($\sim 0.671$nA) while the total current in AP is practically zero ($\sim 0.018$nA) resulting in a large magnitude of TMR (3690\%).

\section{Conclusion}

In conclusion, in this study, we investigated the spin-polarized transport properties of six MnBi$_2$Te$_4$-based AFM-TJ devices,  sandwiched between two silver electrodes. Our studies show that for small devices the I-V characteristics reflect the symmetry of the structural and magnetic configuration of the device. For small devices, the interference due to the confinement effect can favor different spin channels depending on the bias voltage which can result in a non-monotonous behavior of the TMR especially at a small bias voltage. The presence of an intermediate BN layer can significantly enhance the TMR by suppressing particular spin channels for different magnetic configurations. By exploiting this mechanism, we demonstrate that an Ag-MBT-BN-MBT-MBT-MBT-Ag AFM-TJ can exhibit remarkable TMR at specific bias voltages. The TMR value reached up to 3690\%, which could be explained by the spin-polarised transmission channel and  projected local density of states. This study therefore establishes a solid foundation for future research on antiferromagnetic spintronic devices.

\begin{acknowledgments}
This work is supported by the National Natural Science Foundation of China (Grant No. 51671114 and No.U1806219) and the Special Funding in the Project of the Taishan Scholar Construction Engineering. L.Z. acknowledges the Alexander von Humboldt Foundation for funding her postdoctoral research.
\end{acknowledgments}

\appendix

\section{Device construction}

The interfaces were constructed using the QuantumATK interface builder module \cite{stradi2017method}. For the MBT/MBT interface, a 30° rotation angle was applied during alignment, while the angle between the vectors of the created interface supercells was set to 60°. No mean strain was applied to the surfaces during the matching process. For the MBT/Ag interface, a 10.89° rotation angle was used during alignment, with an angle of 60° between the vectors of the created interface supercells. The mean strain applied to the Ag surfaces during matching was 0.78\%. For AA' stacking of BN, the BN layers are rotated by an angle 180°. The mean strain applied to the BN surface during matching was 0.37\%. As for the interface of h-BN/h-BN, it adopts the commonly used stacking mode, namely the antiparallel alignment between adjacent layers, known as AA' stacking. In this stacking arrangement, boron atoms are vertically aligned with nitrogen atoms of neighboring layers. This particular stacking mode offers remarkable thermal and chemical stability, as well as energetically \cite{zhao2021universal}.

\begin{figure}[h!]
\centering
\includegraphics[width=0.5\textwidth]{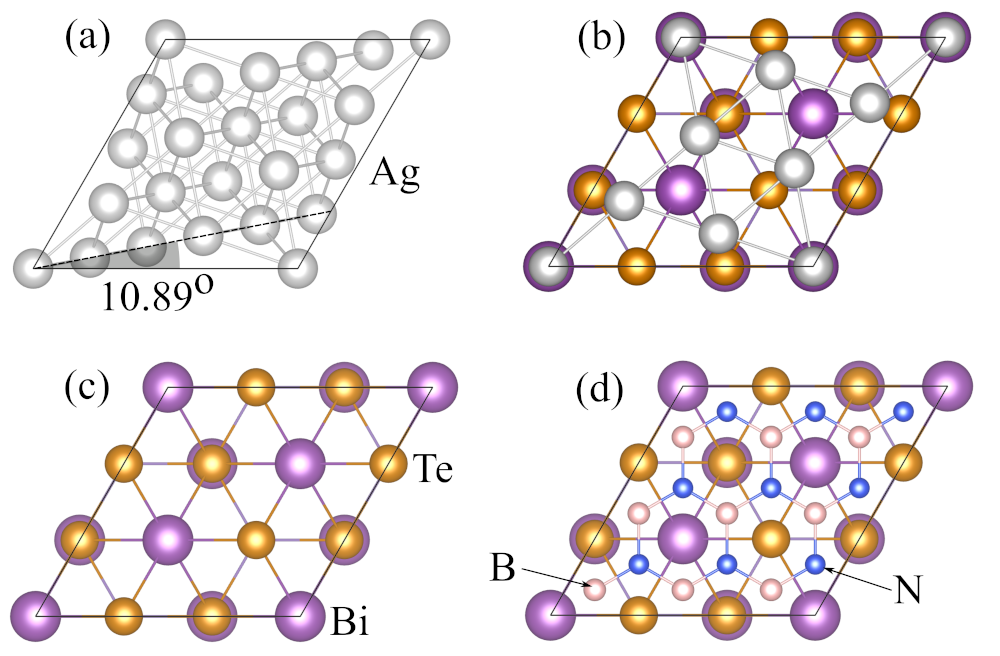}
\caption{Structure of different composition layers used in different AFM-TJ. (a) The Ag layers are rotated by 10.89° to minimize the lattice mismatch with MBT (b). Top view of (c) MBT layer and (d) MBT-BN interface.}
\label{fig:8}
\end{figure}

\section{Effect of strain}
We take the Ag-MBT-BN-MBT-Ag AFM-TJ as an example to examine the influence of strain on the transport properties of the system in this study. We conducted calculations on TMR while varying the in-plane biaxial strain for the Ag-MBT-BN-MBT-Ag AFM-TJ. Under consideration of our previous description that we defined TMR based on non-equilibrium current values instead of equilibrium transmission coefficients \cite{yan2022giant}, we further performed current calculations for different strains under a bias voltage ranging from $-$0.2V to 0.2V. The results are illustrated in Figure 7. From Figure 7, it is evident that regardless of whether compared to compressive stress or tensile stress, the TMR exhibits a higher value when the strain is at zero. However, for both compressive stress and tensile stress, the TMR decreases as the strain intensity increases. Therefore, the above results indicate that when designing AFM-TJs, careful consideration should be given to the influence of strain on the transport properties. The observed decrease in TMR with increasing strain intensity, whether it is compressive or tensile stress, emphasizes the importance of accounting for strain effects in order to optimize the performance of AFM-TJs. By understanding and incorporating the impact of strain, one can make informed decisions in the design process to enhance the transport properties and overall functionality of these devices.
\begin{figure}[ht!]
\centering
\includegraphics[width=0.5\textwidth]{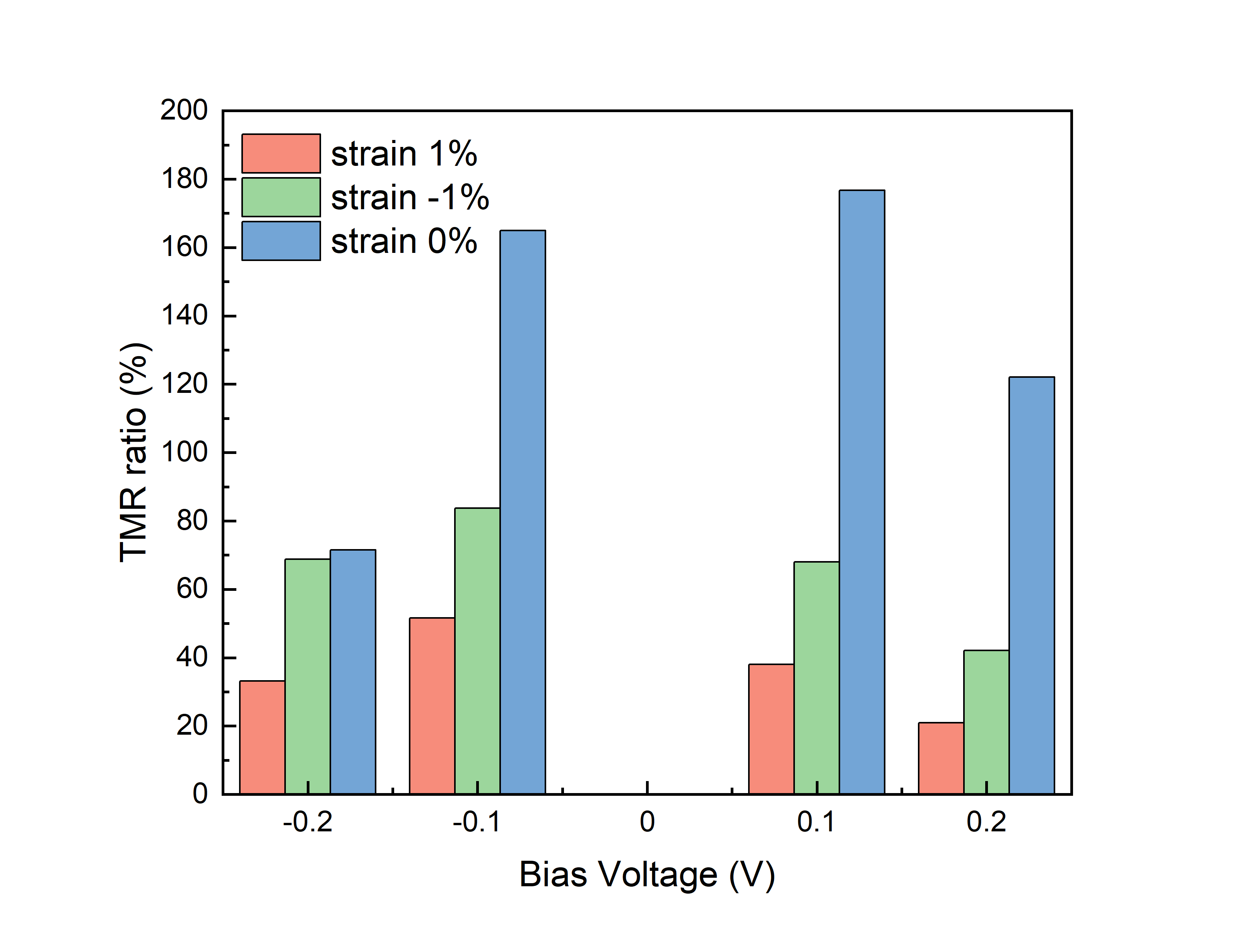}
\caption{The TMR of Ag-MBT-BN-MBT-Ag AFM-TJ in different in-plane biaxial strain.}
\label{fig:9}
\end{figure}

\bibliographystyle{apsrev4-1}
\bibliography{Bibliography}

\end{document}